\documentclass[a4paper,10pt]{article}
\usepackage[utf8]{inputenc}
\usepackage{hyperref}
\usepackage[T1]{fontenc}
\usepackage{amsmath}
\usepackage[symbol]{footmisc}
\usepackage{amsfonts}
\usepackage{setspace}
\usepackage{cite}
\usepackage[dvipsnames]{xcolor}
\usepackage{amssymb}
\usepackage{graphicx}
\usepackage{multirow}
\usepackage[explicit]{titlesec}
\titlespacing\section{0pt}{10pt plus 4pt minus 2pt}{2pt plus 1.5pt minus 1.5pt}
\usepackage{float}
\usepackage{lipsum}
\usepackage{multicol, blindtext, caption}
\usepackage{xcolor}
\usepackage[font=footnotesize,labelfont=bf, skip=10pt]{caption}
\definecolor{azul}{rgb}{0, 0, 0}
\usepackage[left=1.5cm, right=1.5cm, top=2.0cm, bottom=2.5cm]{geometry}
\usepackage{hyperref}

\usepackage[normalem]{ulem}

\usepackage{placeins}
\begin{document}

	\begin{center}
		{\Large \textbf{Schottky barrier height modulation and memristive behavior in a mechanically exfoliated $MoS_2$ flake with multiple lateral electrodes}}\\
		\vspace{5mm}
		Gabriel Caballero$^{1,2,*}$, Ramón Bernardo-Gavito$^{1,3}$, Mariela Menghini$^{1,3,*}$ and Daniel Granados$^{1,3,*}$ \\
		\vspace{3mm}
		$^1$\textit{IMDEA Nanociencia, Cantoblanco, 28049 Madrid, Spain}\\
		$^2$\textit{Facultad Ciencias Físicas, Universidad Complutense de Madrid, 28040 Madrid, Spain}\\
        $^3$\textit{Unidad de Nanomateriales Avanzados, IMDEA-Nanociencia, Unidad Asociada al CSIC por el ICMM, C/ Faraday 9, 28049 Madrid, Spain}
       
	\end{center}

Van de Waals materials possess great potential as core active materials for novel electronics, optoelectronics, and sensing applications. In particular, transition metal dichalcogenides (TMDs) and other low-dimensional materials, such as hexagonal boron nitride  and binary oxides exhibit memristive behavior, making them attractive candidates as building blocks for resistive memories and neuromorphic computing devices. In this work, we investigate two-terminal lateral memristors based on mechanically exfoliated few layer $MoS_2$ flakes. We study the influence of the Schottky barrier height (SBH) as a function of the contact area. We report different SBHs from 60 meV up to 208 meV in the same $MoS_2$ flake and analyze the effect of the SBH in the memristive-like $J-V_{DS}$ curves and the observed non-volatile/volatile resistance switching. Our results highlight the importance of  contact engineering to obtain different memristive properties in $MoS_2$-based devices.  

\vspace{2mm}

$^{*}$ Corresponding authors: gabriel.caballero@imdea.org, mariela.menghini@imdea.org, daniel.granados@imdea.org

\vspace{2mm}
\underline{\textbf{Keywords:}} \hspace{2mm} \textbf{ Memristor, Schottky barrier height, volatile, non-volatile, $\mathbf{MoS_2}$}

\begin{multicols}{2}
\section{Introduction} \label{Intro}

Two-dimensional (2D) crystals of transition metal Dichalcogenides (TMDs) have been studied extensively in recent years due to their wide variety of interesting properties such as, tunable bandgap, mechanical flexibility, high optical absorption and strong spin-orbit coupling, to name a few \cite{Rani2024AdvancementsStrategies}. Since the introduction of the first single-layer molybdenum disulfide ($MoS_2$) field effect transistor (FET) by B. Radisavljevic et al. \cite{Radisavljevic2011Single-layerTransistors}, TMDs were shown to be relevant as single-layer or few-layer semiconductors \cite{Wang2012ElectronicsDichalcogenides} for future nano-electronics and optoelectronics applications due to their relatively high mobility and on-off ratio \cite{Shih2014TuningModulation}, immunity to short channel effects \cite{Liu2012ChannelMOSFETs}, and abrupt switching capabilities \cite{Yoon2011HowBe}. Moreover, $MoS_2$ has proven to be a good candidate as a memristor for possible applications as memory in conventional and non-conventional computing architectures \cite{Fu2023Two-TerminalNetworks, Sangwan2015Gate-tunableMoS2, Sangwan2018Multi-terminalDisulfide, Wells2024MolybdenumProspects}. 

In memristive devices, the resistance state depends on the electrical history of the device. Following a specific voltage cycling process, it is possible to switch the resistance of the device between a low (LRS) and high (HRS) resistance state. Different phenomena in $MoS_2$ devices have been identified as responsible for their memristive behavior, such as defect migration \cite{Jadwiszczak2019MoS2Irradiation}, Schottky barrier modulation \cite{Li2018MoS2Emulation, Fejzaj2026Exfoliated-MoS2Synapses},  or conductive filament formation \cite{Farronato2022MemtransistorMigration, Ran2025UnravelingMicroscopy}.
 Long retention times of approximately 27 hours  and  endurance as large as $10^3$ cycles have been reported in vertical \cite{Li2024Wafer-ScaleSwitching,Tang2022Wafer-scaleComputing} and lateral $MoS_2$ memristor \cite{Lee2020Dual-GatedArray,Sangwan2018Multi-terminalDisulfide}.


Reliable operation of memristors is based on a comprehensive understanding of the interfaces between the active material and electrodes and the manner in which this interface affects the characteristic current density-voltage ($J-V_{DS}$) curves. The contact interface quality can play a relevant role in the control of ion migration, filament formation and charge injection during the switching process and therefore influence the device macroscopic properties such as endurance, operating voltage, reproducibility and stability \cite{Lanza2019RecommendedDevices, Mehonic2024RoadmapTechnologies, Shvetsov2026InterfaceReliability, Braun2026AdvancedMemristors}. An important effect to consider when studying memristors is the formation of a Schottky barrier at the interface due to the mismatch between the electron affinity of the semiconductor and the metal work function. The Schottky barrier height (SBH) is defined as follows:
\begin{equation}
    q\Phi_B = W_{Metal} -\chi_{Semiconductor}
    \label{eq:DefSBH}
\end{equation}

Where $W_{Metal}$ is the metal work function and $\chi_{Semiconductor}$ is the semiconductor electron affinity.

It has been widely demonstrated that the Schottky junction can play an important role in the electronic properties of semiconductor devices \cite{Wei2024AnalyticalPhototransistors}.
Schottky junctions of 2D materials in contact with a certain metal, although fabricated following the same process, often have different $J-V_{DS}$ characteristics due to the dielectric disorder in the 2D materials, which is reflected in different barrier heights \cite{Raja2019DielectricMaterials} and therefore different memristive properties. Different SBH values reported in the literature for different metal/$MoS_2$ structures are shown in Table \ref{tab:sbh}.

\begin{center}
\small
\label{tab:sbh}
\begin{tabular}{c|c}
\hline
Structure & SBH [meV]\\
\hline \hline
\multirow{3}{*}{$Ti/MoS_2$} & 50 \cite{Kim2018SchottkyStates} \\
\cline{2-2}
                            & 95 \cite{Andrews2020ImprovedInterlayer} \\
\cline{2-2}
                            & 250 \cite{Sridevi2021LoweringTransistor}\\
\hline \hline
\multirow{2}{*}{$Au/MoS_2$} & 55 \cite{Kim2018SchottkyStates} \\
\cline{2-2}
                            & 150 \cite{Qiu2015ElectricallyTransistors} \\
\hline \hline
$Ti/TiO_2/MoS_2$ & 107 \cite{Kim2018SchottkyStates}\\
\hline \hline
$Au/TiO_2/MoS_2$ & 212 \cite{Kim2018SchottkyStates} \\
\hline
\end{tabular}
\captionof{table}{Schottky barrier heights (SBH) reported for different metal/$MoS_2$ contacts.}
\end{center}
Memristors can be divided into four different types depending on the combination between non-volatile or volatile resistance switching and whether the switching sequence goes from HRS to LRS or the other way around (see Table S1 in the Supplementary Information). At the same time, in Schotkky barrier-modulated memristors, the switching mechanism can be divided into two categories: in the first one the defects migrate in the same (opposite) direction as the electric field and the SBH at the interface increases (decreases)  while in the second one the defects migrate in the same (opposite) direction of the electric field causing a decrease (increase) in the SBH. In Ref. \cite{Zhou2023Design-DependentSemiconductor} the authors analyze the conditions for the possible combinations of memristor type and switching mechanism that can be stabilized by varying the defect concentration at each contact region.

In this work, we study the current density- voltage characteristics, $J-V_{DS}$ (where, D refers to drain and S to source, similarly to FETs), of multiple lateral symmetric contact devices with different areas fabricated on the same mechanically exfoliated $MoS_2$ flake. We extract the SBH as a function of the contact area for the same  $Au/Ti/MoS_2$ interface and for different voltage polarities. At sufficiently large bias voltage, different memristive (non-volatile and volatile) behavior are observed depending on the contact size. We analyze the symmetry of the memristive $J-V_{DS}$  curves and the resistance switching window as a function of the contact area. The main contribution of this work is to help understanding the influence of interface properties on the macroscopic memristive response of devices based on $MoS_2$.



\begin{figure}[H]
  \small
  \centering
    \includegraphics[width=0.4\textwidth]{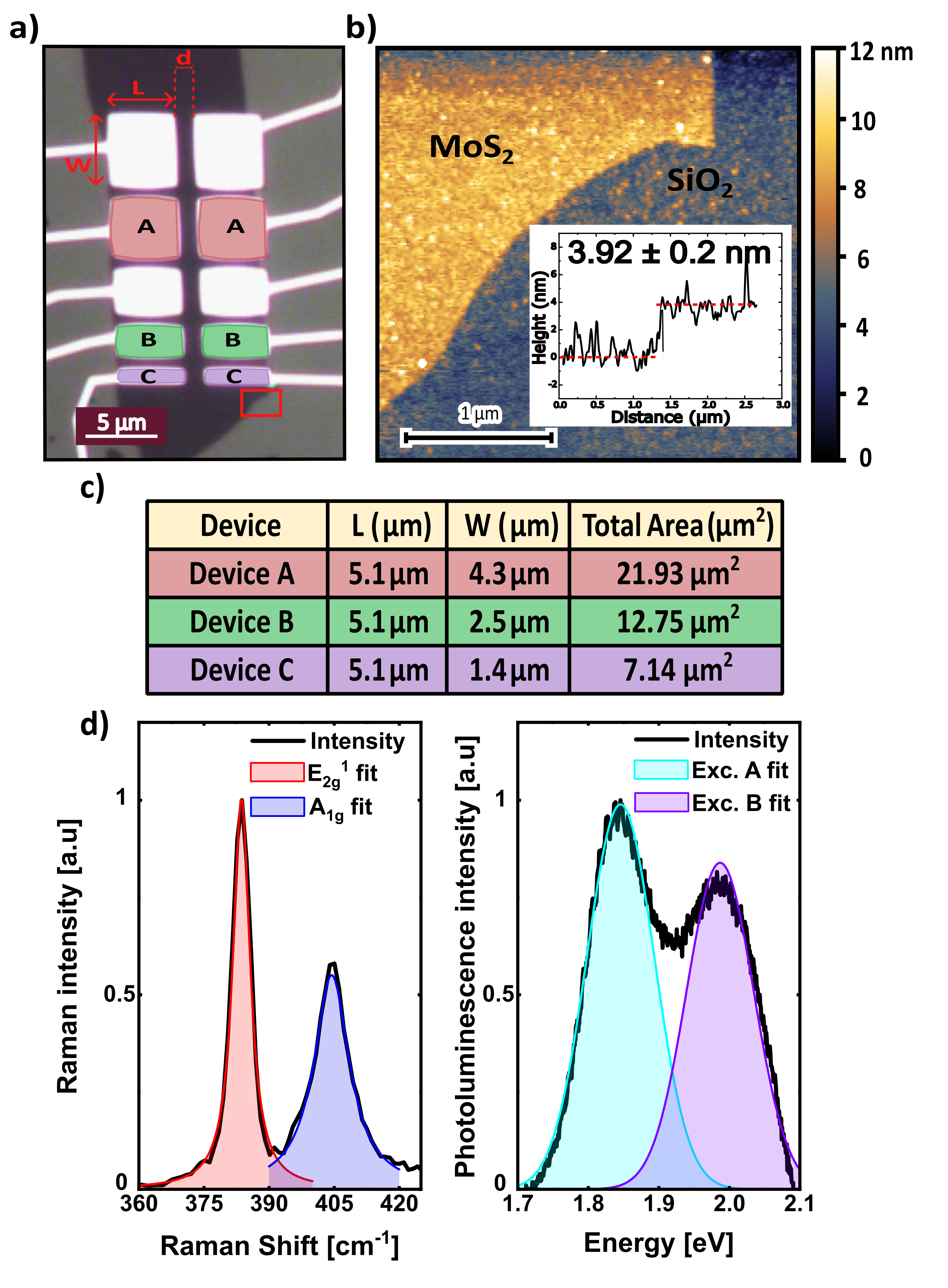}    \caption{a) Optical micrograph of a geometrical defined device with different contact areas. b) Atomic Force Microscopy (AFM) image of the region in a red square in a). Insert: Height of $MoS_{2}$ flake. In Fig. S1 is shown all the AFM images of the three devices. c) Geometrical parameters of the different devices. d) Left: Raman Spectra and Lorentzian fit of $E_{2g}^1$ and $A_{1g}$ $MoS_{2}$ Raman modes at $\lambda = 488 nm$. Right: Photoluminescence Spectra and Gaussian fit of exciton A and B in $MoS_{2}$ at $\lambda = 488 nm$.}
    \label{Fig1}
\end{figure}

\FloatBarrier
\section{Results and discussion} \label{Results}

\begin{figure}[H]
  \centering
    \includegraphics[width=0.4\textwidth]{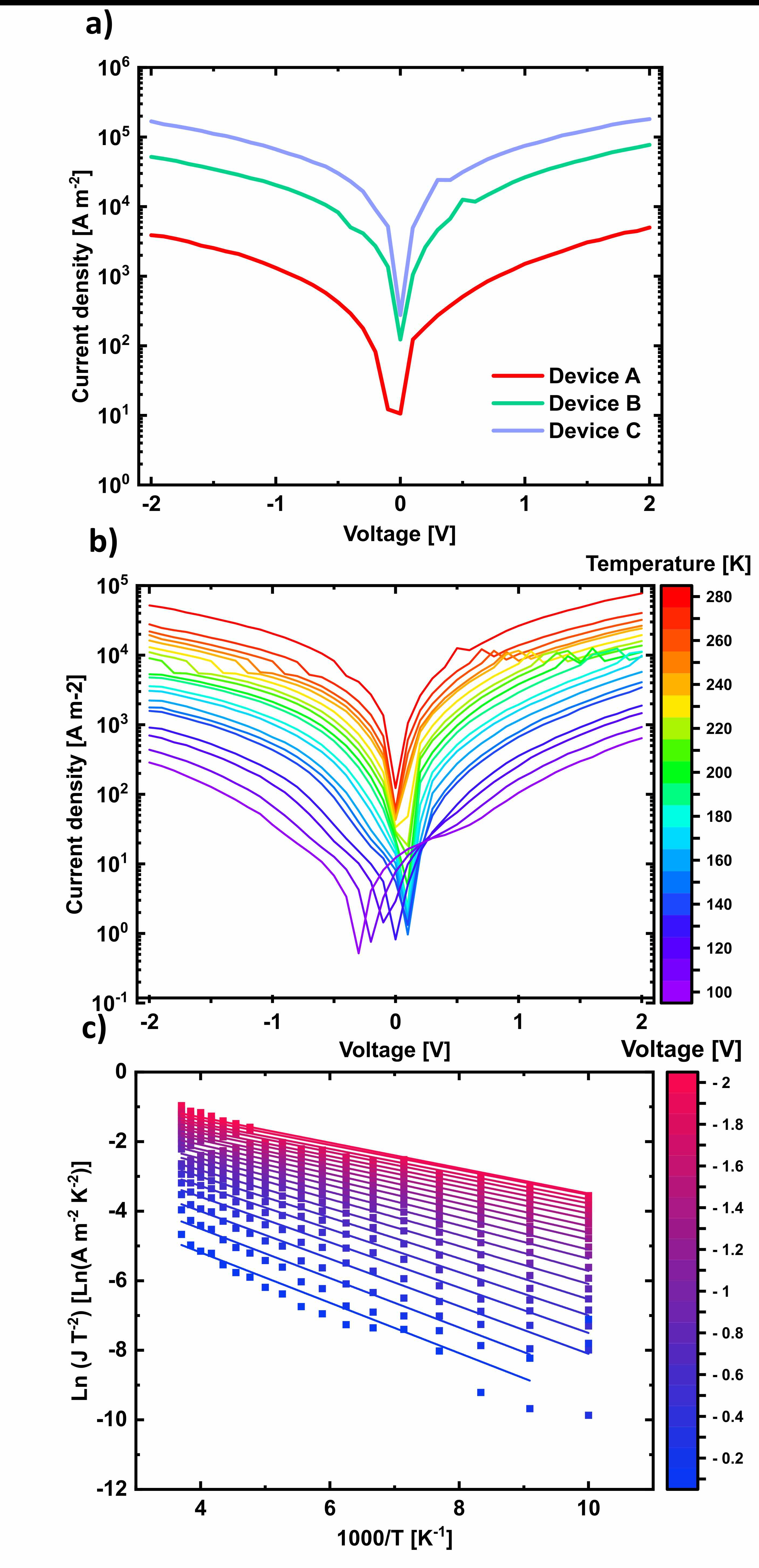}    \caption{a) Current density vs Voltage ($J-V_{DS}$) curves for Devices A, B and C from $V_{DS}=-2V$ to $V_{DS}=+2V$, in steps of $0.1V$. b) Current density vs voltage ($J-V_{DS}$) curves as a function of temperature for Device B. The range of temperatures goes from $T = 280K$ to $T = 100 K$, with $\Delta T = 10 K$. c) Thermionic analysis: $Ln(JT^{-2})$ vs $1000/T$ for Device B and negative voltage bias.}
    \label{Fig2}
\end{figure}

The devices studied consisted on a $MoS_2$ flake obtained by the mechanical exfoliation process with a series of metal electrodes of different sizes deposited on top, see Fig. \ref{Fig1} a. All data presented in this manuscript correspond to a $MoS_2$ flake with a thickness of $\approx 4\ nm$ obtained by Atomic Force Microscopy (AFM) (See insert Fig. \ref{Fig1} b).
To probe the quality of the $MoS_{2}$ flake, Raman and Photoluminescence (PL) spectroscopy were performed at room temperature (Figure \ref {Fig1} d). As expected, both Raman modes, $E_{2g}^1$ and $A_{1g}$ characteristic of $MoS_{2}$ were observed at 383.83 $cm^{-1}$ and 406 $cm^{-1}$, respectively. The PL spectra indicate that Exciton A is located at 1.84 eV and Exciton B at 1.98 eV. The Raman and PL spectra fit correctly to Lorentzian and Gaussian peaks \cite{Li2012QuantitativeSubstrates, Christopher2017LongSpectra}, respectively, confirming the good quality of the crystal. The intensity, position and FWHM obtained from the fits are summarized in Fig. S3. We measured the electrical characteristic properties of five different devices, nevertheless, in this article, we focus on the electrical properties of lateral Devices A, B and C as indicated in Fig. \ref {Fig1} a. For devices A and B, the distance to the next device electrodes is $\approx1\ \mu m$, while device C is located at the edge of the contact array. In these devices, the $MoS_2$ channel length ($d=1.5\mu m$) and the electrode length ($L=5.1\mu m$) are the same, while the electrode width (W) varies from $1.4 \mu m$ to $4.3\mu m$ (the dimensions of the three devices are summarized in Fig. ~\ref{Fig1} c.)
We measured $J-V_{DS}$ characteristics for Devices A, B and C  in a low voltage range in order to determine the relevant transport mechanism. Fig.\ref{Fig2}a shows room temperature $J-V_{DS}$ for the three devices in the voltage range from -2V to 2V. The curves are fairly symmetric with respect to the applied voltage sign and the conductivity increases as the total contact area decreases.
This implies that the current  $I = {J}{A_{contact}}$ does not scale with the contact area, indicating the possible presence of different Schottky barriers in each device at the same bias. It is reasonable to expect that in this type of configuration the $J-V_{DS}$ is mostly determined by the conduction at the $MoS_2$-metal interface.  In a metal-semiconductor junction, the possible electrode-limited conduction mechanisms are: (a)thermionic emission , (b) Fowler-Nordheim tunneling emission and (c) direct tunneling emission \cite{Chiu2014AFilms}.
At high temperatures, we can disregard the tunneling mechanisms \cite{Su2017TemperatureMethod,Son2018Temperature-dependentHeterojunctions}. 
Therefore, to determine the Schottky barrier height we assume the thermionic emission conduction model described as:
\begin{equation}
    J=A^{*}T^2\exp\left[\frac{-q\left(-\Phi_{B}-\sqrt{\frac{qE}{4\pi\epsilon_{r}\epsilon_{0}}}\right )}{k_{B}T}\right]
    \label{eq:Schottky}
\end{equation}
Where $A^{*}$ is the Richardson constant, $T$ is the temperature, $q$ is the electron charge, $q\Phi_{B}$ is the SBH, $\epsilon_{0}$ the vacuum permittivity, $\epsilon_{r}$ the medium permittivity, $E$ is the electric field and $k_{B}$ is the Boltzmann constant.
As an example of the thermionic emission analysis performed in our devices, in Fig. \ref{Fig2} b the $J-V_{DS}$ curves for device B in the temperature range between 280 K and 100 K are shown (the corresponding $J-V_{DS}$ curves for devices A and C can be found in Fig. S4 a and b, respectively). From these characteristics, we calculated the $Ln\left(\frac{J}{T^2}\right)$ for each applied bias voltage. In Fig. \ref{Fig2} c the resulting $Ln\left(\frac{J}{T^2}\right)$ vs $\frac{1000}{T}$ for Device B is shown. The different colors correspond to different bias voltage as indicated in the color bar and the straight lines are the result of a linear fit of the data. According to Eq.\ref{eq:Schottky} the slope of the linear fit can be expressed as: 
\begin{equation}
    \Pi_{Ther} = \alpha*q\Phi_B+\beta*\sqrt{E}
    \label{eq:Slope}
\end{equation}
where $\alpha = \frac{1}{1000k_B}$ and $\beta=q\sqrt{\frac{q}{4\pi\epsilon_0\epsilon_r}}/{1000K_B}$  are constants. 

\begin{figure}[H]
  \centering
    \includegraphics[width=0.8\textwidth]{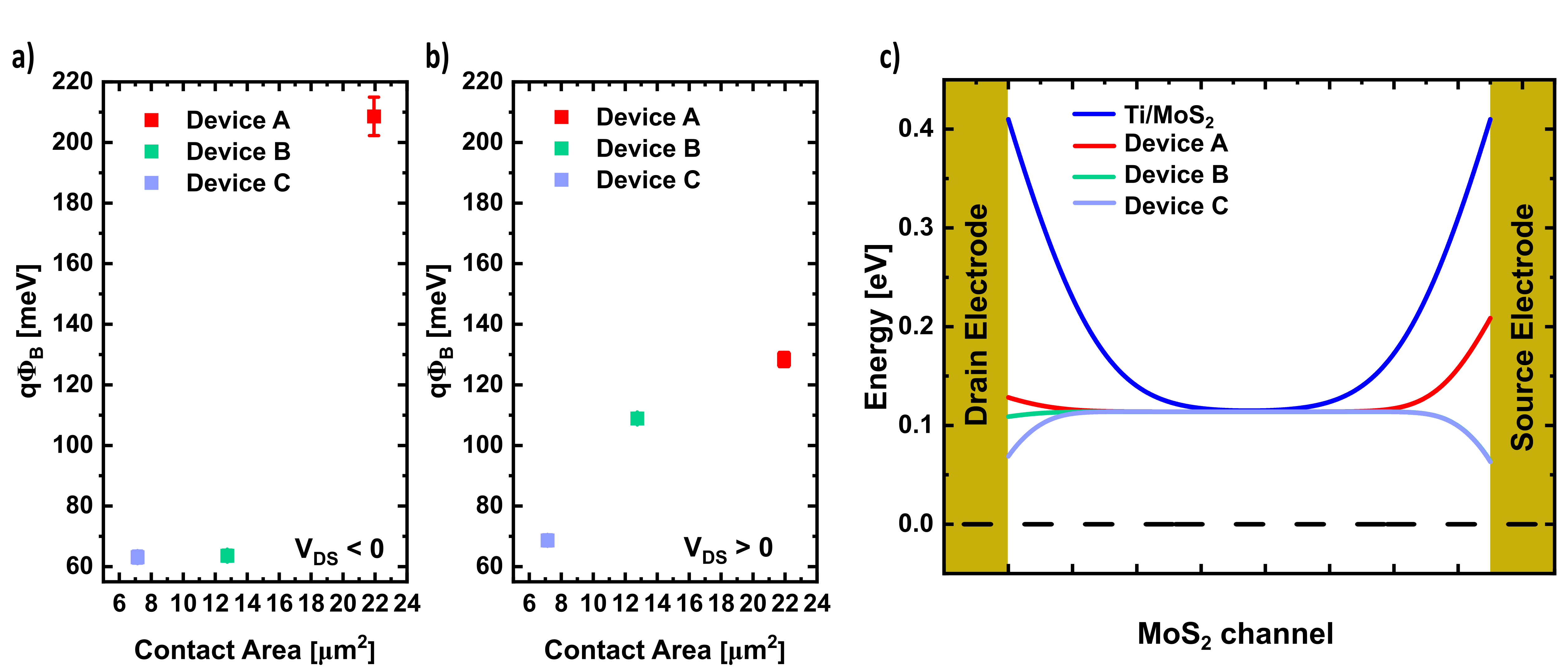}    \caption{Schottky barrier height (SBH) as a function of contact area for negative (a) and for positive (b) voltage bias ($V_{DS}$). The error bars are estimated from the fitting process described in the supplementary material. In all cases, except for Device A for negative bias, the error bar is indistinguishable. (c) Simulation of a Metal Drain - Semiconductor ($MoS_{2}$)- Metal Source  double Schottky barrier geometry for $MoS_{2}$ channel considering the SBHs shown in a) and b). The blue line corresponds to the ideal $Ti/MoS_2$ interface.}
    \label{Fig3}
\end{figure}

Therefore, we can obtain the SBH from the intercept of the plot of $\Pi_{Ther}$ vs $\sqrt{E}$ (see Fig. S6 d).
In this case, i.e. for Device B and negative voltage bias, the obtained value for $q\Phi_B$ is 63.6 meV. We followed the same procedure  for the three different devices and for negative and positive bias voltage (see Supplementary Material,  Figs. S5 a-f and Figs. S6 a-f).
The obtained SBHs for negative and positive bias voltage for the three devices studied are shown in Fig. \ref{Fig3} a and b, respectively. In both cases, the SBH tends to increase with contact area. For $V_{DS} < 0V$ SBH is around 63 meV for the smaller devices and more than a factor of two larger for device A. In contrast, for $V_{DS} >0V$ SBH increases continuously from 68.7 meV up to 128.4 meV as the contact area increases. Scanning Photocurrent Microscopy (SPCM) studies in the same flake are in agreement with a Schottky diode-like behavior (see Fig. S7). In an ideal interface, the SBH depends only on the difference between the metal work function and the semiconductor electron affinity, as mentioned in the introduction \ref{eq:DefSBH}. The variation of SBH with the voltage bias sign and contact area can in principle be due to: a) disorder in the 2D-semiconductor, b) defects or residues from the fabrication process \cite{Cruces2025VolatileDevices, Li2018MoS2Emulation},  and c) Fermi level pinning \cite{Li2024Wafer-ScaleSwitching, Kim2017FermiDichalcogenides}. 

Fig. \ref{Fig3} c, shows the conduction bands calculated at equilibrium solving the 1D Poisson-Boltzmann equation by the finite differences method in a $MoS_2$ and Ti contacts stack (all parameters used in the calculation are summarized in Fig. S8). The blue curve corresponds to the ideal $Ti/MoS_2$ interface, i.e. $q\Phi_B = 413.3\, meV$ while the red, green and purple curves represent the conduction bands considering in the calculations the SBHs values obtained experimentally for devices A, B and C, respectively . In all three devices, the SBHs are smaller than for the ideal interface. The horizontal dotted line corresponds to the Fermi level set at 0 eV. As expected, the conduction band of the device with the highest current density is the closest to the Fermi level at the interface with the metal. 

\begin{figure}[H]
  \centering
    \includegraphics[width=1\textwidth]{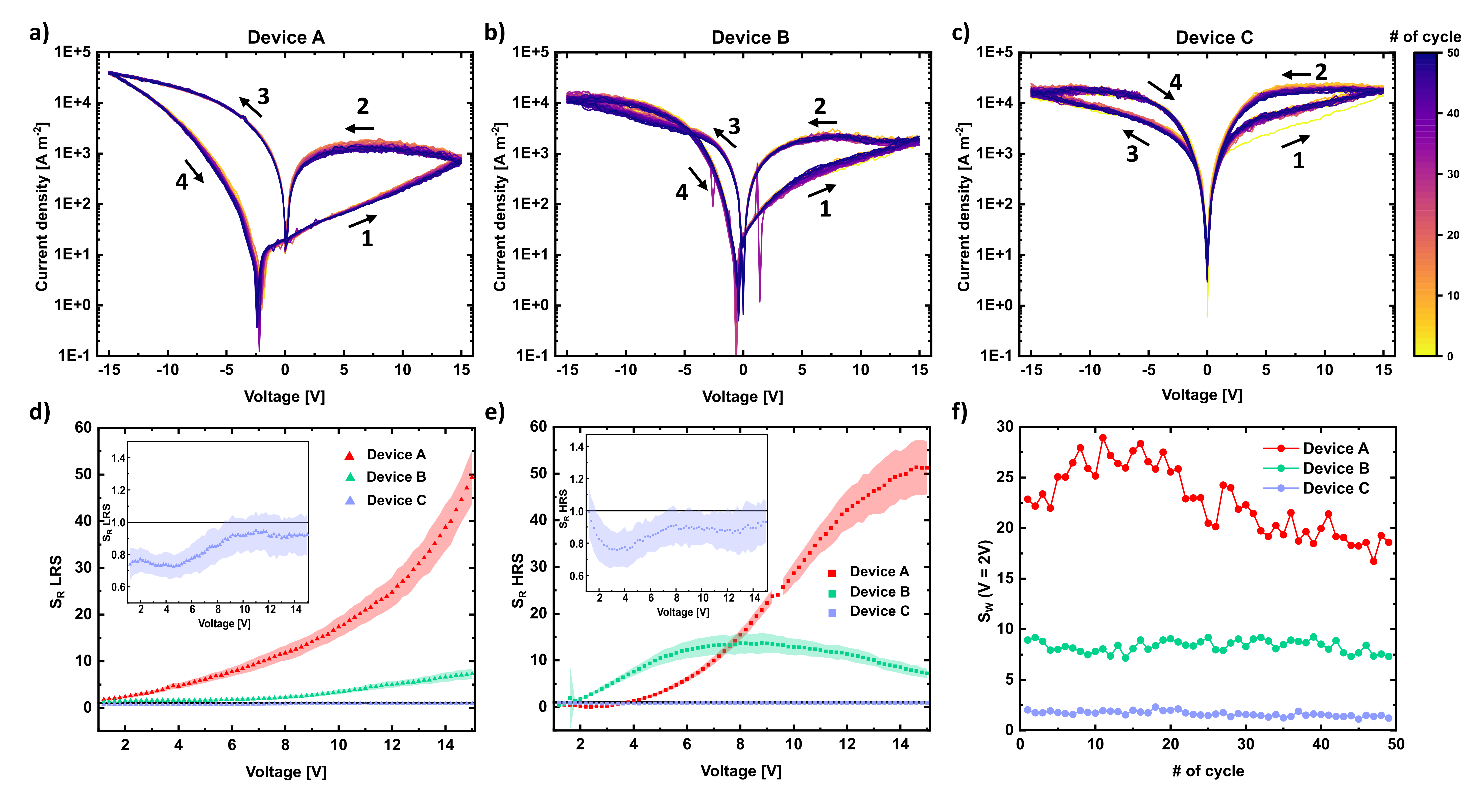}    \caption{ Electrical characterization of $MoS_2$ memristor. a-c) Current density vs voltage ($J-V_{DS}$) curves for 50 consecutive cycles of Devices A (a), B (b) and C (c). Symmetry ratio ($S_R$) for mentioned devices in Low Resistance State (LRS) (d) and High Resistance State (HRS) (e). Insert in Fig. d) and e): $S_R$ for device C in the LRS and HRS, respectively. f) Switching window $S_W$ at $V_{DS}=+2V$ as a function of the number of cycles for all the mentioned devices.The $S_W$ maps for all measured $V_{DS}$ and number of cycles are shown in Fig. S10.}
    \label{Fig4}
\end{figure}

Next, we studied the influence of the contact area on the memristive behavior of the devices. For this, 50 consecutive $J-V_{DS}$ curves were measured in each device, as shown in Fig. \ref{Fig4} a, b and c. The full voltage sweep can be divided in four consecutive sweeps: $1^{st}$ sweep: from 0 V to 15 V , $2^{nd}$ sweep: from 15 V to 0 V, $3^{rd}$ sweep: from 0 V to -15 V and $4^{th}$ sweep: from -15 V to 0 V. A clear hysteresis in the $J-V_{DS}$ curves is observed indicating a memristive behavior. 
Device A goes from a high resistance state (HRS) in the $1^{st}$ sweep to a low resistance state (LRS) in the $2^{nd}$ one (SET). When changing the sign of the voltage ($3^{rd}$ sweep) LRS is preserved, indicating a non-volatile behavior, the device switches back to HRS during the $4^{th}$ sweep (RESET). Endurance tests in Device A could not be performed due to breakdown of the device. Device D  is located next to A in the same flake (see Fig. S2) and presents a similar $J-V_{DS}$ characteristic as Device A. Endurance tests in this device shows a good reproducibility of HRS and LRS up to 500 cycles (see Fig. S9). Therefore, this result also supports some level of non-volatility in these devices. The time scale of the non-volatility could not be assessed as retention tests could not be performed due to breakdown of the devices. Other $MoS_2$ lateral devices fabricated following a similar protocol but on different flakes exhibiting the same $J-V_{DS}$ tracing behavior ($HRS \xrightarrow[]{} LRS \xrightarrow[]{} LRS \xrightarrow[]{} HRS$) have shown retention up to 1 h (not shown).
Devices B and C show the same behavior for positive voltage branch ($1^{st}$ and $2^{nd}$ sweep) as Device A. In contrast to Device A, Device C presents volatile behavior in the negative branch (HRS ($3^{rd}$ sweep) $\longrightarrow$ LRS ($4^{th}$ sweep)). Device B shows an intermediate behavior. During the $3^{rd}$ sweep the current density reaches approximately $10^{-4} A/m^2$ indicating the stabilization of a LRS.
\end{multicols} 
\vspace{0.5mm}

\vspace{1mm}
\begin{multicols}{2}
However, when voltage is swept back to zero, the current density is slightly higher than the one during the  $3^{rd}$ sweep and the hysteresis is significantly smaller in the range $V_{DS}=[-15 V,-5V]$. Then, in the voltage range  $V_{DS}=[-5V,-0V]$ the device shows non-volatile behavior as in the case of Device A.

One might assume that the location of the device on the flake and whether there are other devices at the sides can influence the memristive characteristics. However, as mentioned above, non-volatile-like $J-V_{DS}$ curves are also observed in a large device located at the edge of the flake (Device D, see Fig. S9 a). Moreover, volatile behavior has been observed on isolated and non-isolated devices on different flakes (see Fig. S11). Therefore, no clear influence of neighboring devices in memristive properties can be assessed. This is supported by SPCM experiments showing that the photocurrent is clearly concentrated only within the active electrode (see Fig. S7).
According to the literature \cite{Zhou2023Design-DependentSemiconductor}, this memristive behavior in $MoS_{2}$ devices can be explained by SBH modulation and current rectification due to the presence of defects. Non-volatile resistive switching behavior could be induced by defects at the interface between Ti/Au electrodes and the $MoS_{2}$ surface (see Fig. S12 a-d), resulting in a higher SBH in Devices A and B. Interfacial defects can also be present in devices showing a $J-V_{DS}$ curve compatible with a volatile response. However, the defect density or trap depth could be low enough so that a non-volatile state can not be stabilized and thermionic emission and SBH modulation are the main responsible for charge transport at the interfaces (see Fig. S12 e-h).
 
Capacitive effects in the $J-V_{DS}$ curves of Devices A and B are evidenced as a finite current at zero voltage. The effect is more prominent in Device A than in B. Capacitive effects in TMDs have been associated with trapped states at the interface originated by adsorbates or defects in the material. \cite{Lee2020Dual-GatedArray, Sangwan2018Multi-terminalDisulfide, Guo2024MoS2Memcapacitor}. A larger capacitive effect in Device A is consistent with the proposed larger number of interfacial defects in this case. In any case, these effects do not seem to affect the order in which the $J-V_{DS}$ is recorded.

Considering the (non-)volatility observed in the $J-V_{DS}$ curves, our devices belong to type 1 ($HRS\xrightarrow[]{}LRS$ and non-volatile behavior) or type 3 ($HRS\xrightarrow[]{}LRS$ and volatile behavior) memristors. According to the proposed phase diagram (Fig. 2 in Ref. \cite{Zhou2023Design-DependentSemiconductor}) and the observation of a reduction of SBH, our devices belong to category 2, where the defects migrate in the same direction as the electric field. For this category, non-volatility is expected when one electrode presents higher defect concentration than the other one, while a volatile behavior is observed when both electrodes have similar defect concentration.
The $J-V_{DS}$ characteristics in Device C are symmetric (positive and negative branches are equivalent), however Devices A and B show an asymmetry between positive and negative voltage sweeps. In order to quantify this behavior, we define a symmetry ratio ($S_R$) as: 
\begin{equation}
    S_R^{LRS} =\frac{I^{LRS}(V_{-})}{I^{LRS}(V_{+})}    ;\,\, \,\,    S_R^{HRS} =\frac{I^{HRS}(V_{-})}{I^{HRS}(V_{+})}
    \label{eq:Symmetry}
\end{equation}
Where $+/-$ indicate the positive/negative voltage branch, respectively. In Fig. \ref{Fig4} d and e, the $S_R$ in the voltage range from 0 to 15 V for LRS and HRS, respectively, are shown. In Device C, both LRS and HRS  symmetry ratios are $\approx 1$ for the whole voltage range with a slight deviation of around 0.3 for voltages below $8V$ (see inserts in Fig. \ref{Fig4} d and e). For Device A , $S_R$ increases monotonically with voltage reaching $\approx50$ for $15V$. In contrast, for Device B, $S_R^{LRS}$ increases continuously, while $S_R^{HRS}$ increases for voltages up to $\approx8V$ and decreases slightly for larger voltages.

Furthermore, we observe a remarkable reproducibility of $J-V_{DS}$ in the three devices studied for 50 consecutive voltage cycles independent of their symmetric character. To study the stability and reproducibility of the memristive states we defined the switching window ($S_W$) as:
\begin{equation}
    S_W =\frac{I_{+}^{LRS}}{I_{+}^{HRS}}
    \label{eq:Switching}
\end{equation}
The $S_W$ for all the devices at $V_{DS}=+2V$ is shown in Fig. \ref{Fig4} f. For devices A and B we can observe one order of magnitude difference between both states (LRS and HRS). For Device C, the $S_W$ is significantly smaller. However, for all  devices, the $S_W$ is quite stable as a function of the number of cycles. The deviation of $S_W$ is around 12\%, 7\% and  15\% for devices A, B and C respectively, indicating good stability of the devices. Therefore, it seems that thermal and/or time dependent effects do not play an important role in the electrical response. 

Our results indicate that defect traps and defect migration are relevant in the electrical response of lateral $MoS_2$ devices. A contact area dependent SBH can be the result of the combination of disorder in the 2D material and possible extrinsic defects. We can´t disregard an inhomogeneous distribution of defects due to the exfoliation process. Therefore, for contacts with larger areas we can expect a larger defect concentration difference between electrodes, and a non-volatile response according to the expected switching mechanism of SBH modulated memristor \cite{Zhou2023Design-DependentSemiconductor}. 
Although the contribution of other transport mechanisms at high voltages can´t be disregarded, the observed non-volatility seems to be related to a larger SBH in devices with larger contact area.
Therefore, the different SBHs as a function of the contact area and the possible different defect concentration between electrodes seem to be the main reason for the observed changes in the memristive behavior in devices fabricated on the same flake. 

\section{Conclusions}
We have studied the electrical properties of three lateral $MoS_2$ devices in a single flake with different contact areas ranging between 7 and 22 $\mu m^2$. 
 Our results show a decrease of $q\Phi_{B}$ at the interface between $MoS_2$ and Ti compared to the theoretical value. This reduction can be due to intrinsic defects in the $MoS_2$ crystal, defects created during the exfoliation process (PDMS residues, strain, etc) and the fabrication process (resist residues, changes in doping, etc). 
For large voltages we observe a hysteresis in the $J-V_{DS}$ curves indicating a memristive-like behavior with a reasonable resistance switching stability for at least 50 cycles.
On one hand, non-crossing $J-V_{DS}$ in larger devices indicate non-volatility confirmed by endurance cycles of at least 500 SET/RESET pulses.  On the other hand, smaller devices show volatile response similar to other devices fabricated on different flakes following the same protocol. The memristive characteristics of our devices are in line with other $MoS_2$ devices from the literature, with some reported devices performing better and others worse (see Table S2 in SI).
The observation of both non-volatile and volatile resistance switching in the same flake could be attributed to an inhomogeneous concentration of defects along the flake.
 The larger devices show a non-volatile behavior likely corresponding to a larger difference in defect concentration between electrodes while the volatile behavior in the smaller device points to a similar defect concentration in  both electrodes. This is supported by an observation or a larger SBH in the larger devices as compared with the smaller ones. 
The SPCM maps and different memristive behavior in the devices at the edges of the flake indicate that the proximity of other electrodes does not influence the memristive characteristics. 
Our work highlights the relevance of electrode engineering and interfacial effects on the memristive characteristics of devices based on van der Waals materials. A deeper and detailed study of the microscopic properties of the interface is needed to tune the interfacial effect in this type of memristors.  By combining volatile and non-volatile behavior, interfacial engineering holds great potential  for developing electronic neurons and synapses based on the same material for neuromorphic computing.

\end{multicols} 
\vspace{1mm}

\vspace{1mm}
\begin{multicols}{2}

\section{Experimental methods} \label{Exp}
\subsection{Device fabrication} \label{Device}
The devices are fabricated on a 290 nm-$SiO_2$/Si(N++) wafer. First, we patterned $Ti(5 nm)/ Au (60 nm)$ large electrode areas and a numeric matrix used to facilitate the localization of $MoS_2$ exfoliated flakes. The substrate is cleaned in a Tepla $O_2$ plasma at 450W and 150ml/min during 2 minutes to clean the surface. The substrate is heated at $125^\circ$ C in a hot plate during 15-20 min to desorb $H_2O$ molecules and improved the exfoliation process. Natural bulk $MoS_2$ crystal is used to obtain $MoS_2$ flakes using the conventional mechanical exfoliation process. We transfer thin $MoS_2$ flakes from a Nitto tape onto a Poly(dimethylsiloxane) film that is then put on top of the cleaned substrate and heated at 125ºC during 15 minutes. We used a Leika microscope zoom $X100$ to find thin and large area $MoS_2$ flakes using the numeric matrix. 

We spin-coated a Poly(methyl 2-methylpropenoate) (PMMA) on top of the substrate and used an Auriga Zeiss electron beam lithography to make a first step of lithography and design the electrodes on top of the $MoS_2$ flake. By e-beam evaporator (Angstrom Engineering) we evaporated $Ti (5nm)/ Au (50 nm)$ and after the evaporation a conventional lift-off process in Acetone and 2-propanol is done. A second optical lithography process (Heidelberg laser writer DVL66FSby ) followed by another thermal evaporation of $Ti(5 nm)/ Au (80 nm)$  was done to connect the first set of electrodes defined by the e-beam lithography with the main pre-patterned ones. 

\subsection{Electrical characterization} \label{Electrical}
The electrical characterization at low temperature was done in an optical Oxford cryostat (lowest available temperature was $4K$) and a Keithley 2450 SourceMeter was used in 2-wire configuration to carry out I-V curves sweeping the voltage from $V_{DS}=-2 V$ to $V_{DS}=+2 V$ with $\Delta V = 0.1 V$ every 10 K. 

The memristive $J-V_{DS}$ curves were done in a probe station using a Keithley 2614B Sourcemeter and by programming a sweep from $0 V \longrightarrow 15 V \longrightarrow 0V \longrightarrow -15V \longrightarrow 0V$. 

Scanning Photocurrent Microscopy (SPCM) was done using a closed-cycle cryostat (Attocube AttoDRY 800) with a coarse XYZ piezoelectric micro positioner (1xAttocube ANPz102/RES/LT) and a Keithley 2450 at 2-wire configuration to apply the voltage bias. The illumination source used to measure the optical signal is a Modu-Laser Stellar Pro Ar-ion laser (CW, 488nm and $0.2 mW$). A serial resistance of $100K\Omega$ was used to monitor the current across the $MoS_2$ channel as a function of the laser position.

\subsection{Optical characterization}\label{Optical}
Raman and photoluminescence spectroscopy were done in a closed-cycle cryostat (Attocube AttoDRY 800) with a coarse XYZ piezoelectric micro positioner. We have optical access to the device  through optical windows and a microscope objective (Attocube LT-APO/VIS/0.82) located inside the cryostat. The illumination source used to measured the optical signal is a Modu-Laser Stellar Pro Ar-ion laser (CW, 488nm) and we used power of 1 mW. A spectrometer (Andor technology, SR-500i-D1) with an indexed triple grating turret was used to measure the Raman and Photoluminescence signal.
\subsection{Microscopy Characterization}\label{Microscopy}
Atomic force microscopy (AFM) measurements were carried out in a commercial AFM system (Ntegra Prima, NT-MDT) in semicontact (dynamic) mode using scanning-by-sample in ambient conditions. Rectangular silicon cantilevers $HQ:NSC15/Al BS$ were used with a tip radius of 8 nm. Their nominal spring constant is 40 N/m and the resonance frequency is around 325 kHz.
Electric force microscopy (EFM) measurements were carried out in a commercial AFM system (Ntegra Prima, NT-MDT) in semicontact (dynamic) mode using scanning-by-sample configuration in ambient conditions. Rectangular silicon cantilevers $SCM_PIT-V2$ were used with a tip radius of 8 nm. Their nominal spring constant is 3 N/m and its resonance frequency is around 75 kHz.
\subsection{Simulations}\label{Simulations}
Band diagram simulations were done modifying the open access code "eq\_band\_diagram" done by Steve Byrnes written in $Python/NumPy$ (https://pypi.org/project/eq\_band\_diagram/) that calculates equilibrium band structures of planar multilayer semiconductor structures solving the "Poisson\-\ Boltzmann\ equation" via finite differences.

\section*{Acknowledgements} \label{Ackowledge}
We thank Enrique Miranda, Jordi Suñe, Fernando Aguirre and Ignacio Figueruelo-Campanero for valuable discussions and Manuel R. Osorio and Fernando J. Urbanos for their help with clean-room fabrication process of the devices. 
This work was supported by the European project MEMQuD, 20FUN06. EMPIR 20FUN06 MEMQuD project has received
funding from the European Metrology Programme for Innovation and Research (EMPIR) cofinanced by the participating states and from the European Union’s Horizon 2020 research and innovation programme. We acknowledge support from the “Severo Ochoa” Programme for Centres of Excellence in R$\&$D (Grants SEV-2016-0686 and CEX2020-001039-S), Spanish Ministry for Science and Innovation (MCIN) PID2021-122980OB-C52 (ECoSOx-ECLIPSE) AEI PID2024-157112OB-C53 (HYBRID-OS: HYPERFAN) and from Comunidad de Madrid  TEC-2024/TEC-380 (Mag4TIC-CM). G. C. holds an PIPF fellowship from Comunidad de Madrid (PIPF-2022/TEC-24324). PID2022-137779OB-C42 funded by the Spanish MCIN/AEI/10.13039/501100011033, and MAD4SPACE-TEC-2024/TEC-182 from the Comunidad de Madrid.

\section*{Data availability statement} \label{Data}
\noindent
The data supporting the findings of this study are available upon reasonable request.

\noindent

\bibliographystyle{unsrt}
\bibliography{References}

\end{multicols}
\end{document}